\documentclass[aps,prl,twocolumn,superscriptaddress,10pt]{revtex4-1}

\usepackage{graphicx}
\usepackage{amsmath}

\usepackage{color}

\begin{document}


\title{Unifying the low-temperature photoluminescence spectra of carbon nanotubes: the role of acoustic phonon confinement.}

\author{F. Vialla}
\author{Y. Chassagneux}
\email{yannick.chassagneux@lpa.ens.fr}
\author{R. Ferreira}
\affiliation{Laboratoire Pierre Aigrain, \'Ecole Normale Sup\'erieure, CNRS (UMR 8551), Universit\'e Pierre et Marie Curie, Universit\'e Paris Diderot, 24, rue Lhomond, F-75005 Paris, France}

\author{C. Roquelet}
\affiliation{Laboratoire Aim\'e Cotton, \'Ecole Normale Sup\'erieure de Cachan, Universit\'e Paris Sud, CNRS (UPR3321), F-91405 Orsay, France}

\author{C. Diederichs}
\affiliation{Laboratoire Pierre Aigrain, \'Ecole Normale Sup\'erieure, CNRS (UMR 8551), Universit\'e Pierre et Marie Curie, Universit\'e Paris Diderot, 24, rue Lhomond, F-75005 Paris, France}

\author{G. Cassabois}
\affiliation{Laboratoire Charles Coulomb, Universit\'e de Montpellier, CNRS (UMR5221), F-34095 Montpellier, France}

\author{Ph. Roussignol}
\affiliation{Laboratoire Pierre Aigrain, \'Ecole Normale Sup\'erieure, CNRS (UMR 8551), Universit\'e Pierre et Marie Curie, Universit\'e Paris Diderot, 24, rue Lhomond, F-75005 Paris, France}

\author{J.S. Lauret}
\affiliation{Laboratoire Aim\'e Cotton, \'Ecole Normale Sup\'erieure de Cachan, Universit\'e Paris Sud, CNRS (UPR3321), F-91405 Orsay, France}

\author{C. Voisin}
\affiliation{Laboratoire Pierre Aigrain, \'Ecole Normale Sup\'erieure, CNRS (UMR 8551), Universit\'e Pierre et Marie Curie, Universit\'e Paris Diderot, 24, rue Lhomond, F-75005 Paris, France}


\date{\today}

\begin{abstract}

 At low temperature the photoluminescence of single-wall carbon nanotubes show a large variety of spectral profiles ranging from ultra narrow lines in suspended nanotubes to broad and asymmetrical line-shapes that puzzle the current interpretation in terms of exciton-phonon coupling. Here, we present a complete set of photoluminescence profiles in matrix embedded nanotubes including unprecedented narrow emission lines. We demonstrate that the diversity of the low-temperature luminescence profiles in nanotubes originates in tiny modifications of their low-energy acoustic phonon modes. When low energy modes are locally suppressed, a sharp photoluminescence line as narrow as 0.7 meV is restored. Furthermore, multi-peak luminescence profiles with specific temperature dependence show the presence of confined phonon modes. 

\end{abstract}


\maketitle



Single-Wall Carbon Nanotubes (SWNTs) are fascinating nanostructures with unique properties that make them promising for a number of opto-electronic applications. The understanding of their basic photo-physical properties has been the focus of an intense research effort with major leaps such as the measurement of an exceptional exciton binding energy \cite{Wang2005}, the detection of trions and bi-excitons \cite{Matsunaga2011,Colombier2012} or the observation of photon anti-bunching due to the localization of excitons \cite{Hogele2008}. 
In contrast, the understanding of the low temperature photoluminescence (PL) spectra of carbon nanotubes remains a pending issue, with apparently inconsistent results reported in the literature. 
For instance, the low temperature luminescence profiles of SWNTs range from broad (more than 10 meV) asymmetrical lines to narrow (less than 40 $\mu$eV) and symmetrical lines \cite{Hofmann2013, Sarpkaya2013, Galland2008, Lefebvre2004, Htoon2004}. 

The general framework of the coupling of a localized exciton to an acoustic phonon bath was developed by Krummheuer \textit{et al.} \cite{Krummheuer2002}. In the case of nanotubes, Galland et al. \cite{Galland2008} showed that this model predicts intrinsically broad and asymmetrical line-shapes. This so-called Ohmic model accounts for a limited set of luminescence profiles but fails to explain the variety of profiles reported in the literature. 
In this model, the red shifted (resp. blue shifted) wing of the PL line corresponds to the radiative recombination of the exciton assisted by the emission (resp. absorption) of an acoustic phonon. 
In contrast to the case of epitaxial quantum dots embedded in a 3 dimensional (3D) phonon bath where a sharp purely electronic Zero Phonon Line (ZPL) accompanied by weak phonon side-bands is observed \cite{Besombes2001,Favero2003}, a strong prediction of this model in the case of a 1D phonon bath is the complete merging of the ZPL into the phonon wings \cite{supmat}.
 Thus, for a 1D phonon 
bath, the dominant optical process near the electronic resonance is the simultaneous emission of a photon and a phonon. As a consequence, the photoluminescence line-shape is driven by this non perturbative coupling to long wavelength acoustic phonons rather than by the eigen properties of the electronic two-level system. 
This peculiar exciton-phonon coupling provides a new means to probe the coupling between exciton and low-energy acoustic phonons with an unprecedented sensitivity.

In this Letter, we present a complete set of PL spectra spanning a large variety of profiles and line-widths. We show that, due to phonon confinement, the coupling of excitons to acoustic phonons can deviate strongly from the Ohmic regime especially for phonon energies lower than a few meV. This has drastic effects on the photoluminescence spectra recovering a sharp ZPL. 
Phonon confinement allows us to explain all the observed spectra both for matrix embedded and suspended SWNTs, provided that the much longer dephasing time of suspended nanotubes is properly taken into account.

 The PL spectra were recorded on individual SWNTs using a homemade confocal microscope.
The surfactant (Sodium Cholate) embedded CoMoCat SWNTs were deposited onto the flat surface of a high refractive index solid immersion lens (SIL) to improve the excitation and collection efficiencies. Prior to SWNTs deposition, the SIL was functionalized with a poly-L-lysine layer to increase the linkage of the micelles to the substrate. 
The sample was placed on the cold finger of an optical helium cryostat.

Optical excitation was performed with a cw Ti:Sapphire laser with intensities of 1kW/cm$^2$ or below. 
Exposure times of typically one minute were used.

\begin{figure*}
\includegraphics[width=17cm]{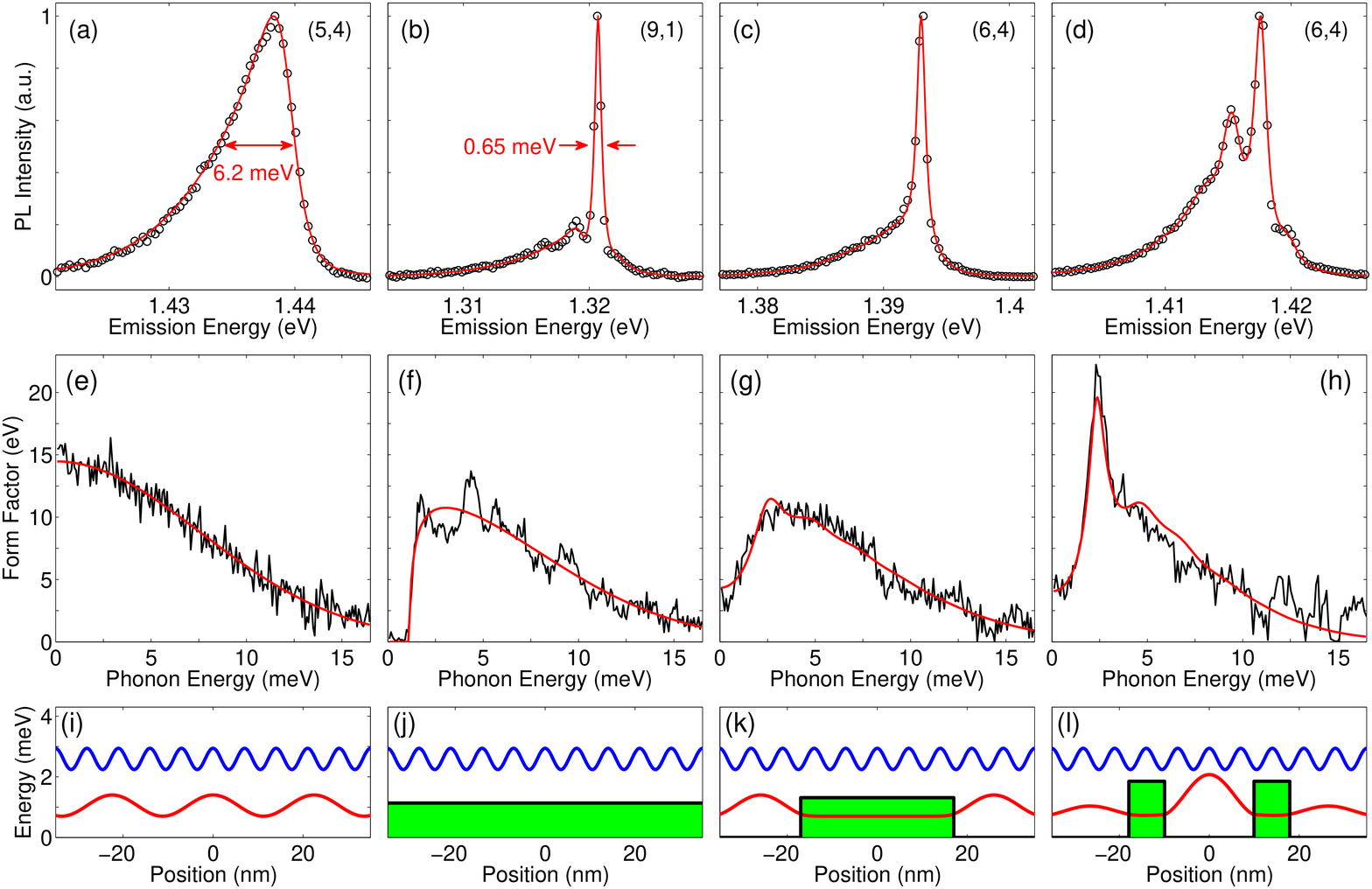}
\caption{(a) to (d) PL spectra of individual SWNTs recorded at 10K (open dots). Chiral species are indicated in parenthesis. 
The red lines are simulated spectra using the fit of the form factor of panels (e) - (h). (e) to (h) Form factors (see text) extracted from the corresponding spectra. 
The continuous lines are fits to the acoustic barrier model \cite{supmat}. (i)-(l) Schematic representation of two exemplary local phonon modes ($\hbar\omega=1$~meV (red line) and $\hbar\omega=2.5$~meV (blue line)) resulting from the presence of acoustic barriers (depicted in green). 
For the spectrum simulation, the exciton is supposed to be located at $z=0$.\label{fig1}}
\end{figure*}

Fig.~\ref{fig1} (a)-(d) show the typical low-temperature PL profiles observed in our samples. The first one (a) consists of an asymmetrical profile and is similar to those reported by Galland \textit{et al.} \cite{Galland2008}. The second one (b) exhibits a sharp peak onto an asymmetrical pedestal. 
Note that this kind of profile is closer to the ones observed for suspended SWNTs \cite{Hofmann2013,Sarpkaya2013}. 
We also found a continuous set of spectra (such as in Fig.~\ref{fig1}~(c)) bridging the narrow and the broad ones. 
Finally, the last example (d) shows a multi-peak feature onto an asymmetrical pedestal. 

\begin{figure}
\includegraphics[width=8.6cm]{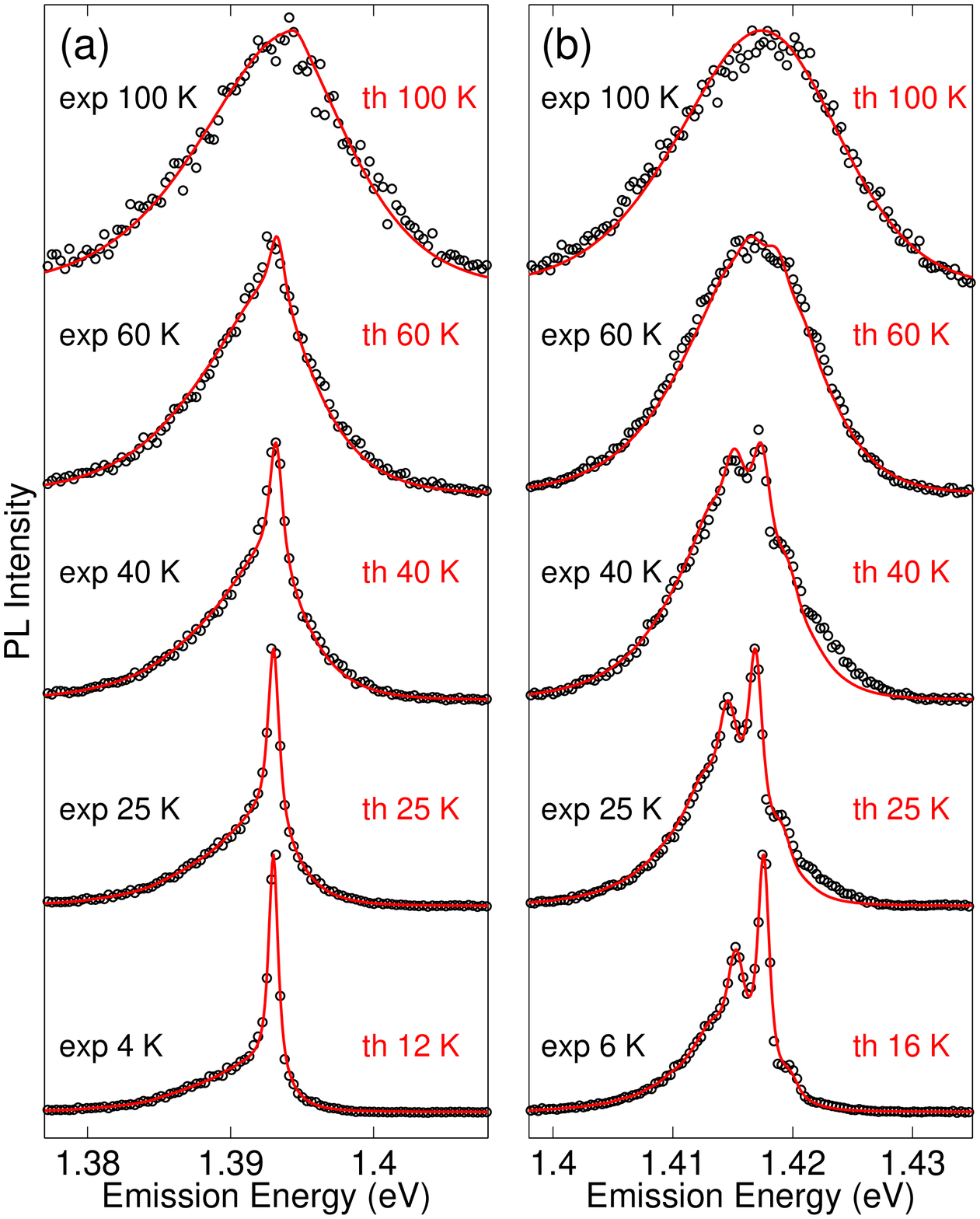}
\caption{Evolution of the normalized PL spectra (vertically shifted  for clarity) as a function of temperature (black open dots).  The red lines are the calculated spectra from the extracted form factor at the lowest temperature (Fig.~\ref{fig1}~(e) to (h)). The measured base plate temperature is indicated together with the effective temperature used for the fitting. \label{figexperiment}}
\end{figure}

In fact, we note that the pedestal (Fig.~\ref{fig1}~(b)-(d)) has -in all spectra- a shape and a width (a few meV) comparable to that of the asymmetrical spectrum (Fig.~\ref{fig1}~(a)). The temperature dependence of the PL line shape of tubes (c) and (d) is given in Fig.~\ref{figexperiment} and shows striking similarities~: when raising the temperature, the blue wing tends to increase leading to a symmetrized profile. 
Importantly, we found that the relative height of the red and blue side peaks in Fig. 1 (d) fits to a Boltzmann factor with an activation energy of 2.5~meV equal to the splitting between the peaks \cite{supmat}. This rules out that these peaks would arise from spectral diffusion. 
Actually, spectral diffusion would not lead to such temperature dependence since there is no simple relationship between the trapping energy of a charge and the magnitude of the related Stark shift \cite{supmat}. 
More generally, we managed to fit the relative weight of the blue and red 
wings to a Boltzmann factor for all spectra, showing that these wings are related to phonon assisted processes \cite{supmat}. Thus, we speculate that all the PL profiles originate from the coupling of excitons to phonons.

In order to extract the coupling between the exciton and acoustic phonons, we used the calculations developed for semiconductor quantum dots in \cite{Krummheuer2002} and adapted for SWNTs in \cite{Galland2008}. We assume that excitons couple only to the 1D acoustic phonons of the nanotube due to the strong sound velocity mismatch between the nanotube ($v \simeq$ 2$\times 10^4$~m/s) and the substrate ($\simeq 3\times 10^3$~m/s). 
In carbon nanotubes, three acoustic phonon branches are coupled to the exciton: the stretching, twisting and radial breathing (RBM) modes \cite{Suzuura2002, Pennington2005}. The acoustic phonons couple to excitons through the difference of the off-diagonal terms of the deformation potential for electrons and holes $D=D_e-D_h$. Due to the complementary dependences with the chiral angle, the different branches can be merged into an effective acoustic mode \cite{Nguyen2011a}.
The exciton-phonon coupling matrix element is given by \cite{Krummheuer2002}: 
\begin{equation}
\hbar g (\omega)=\sqrt{\frac{\hbar\omega}{2\rho L v^2}}  F(\omega)
\label{eq:coupling}
\end{equation}
$L$ is  the nanotube length, $\rho$ is the linear mass density, $v$ is the sound velocity,  $\hbar\omega$ is the energy of the phonon mode and $F(\omega)$ is a form factor. In the case of a perfectly 1D acoustic mode, one has $F=F_0$ (Ohmic coupling) with :
\begin{equation}
F_0(\omega)=D \int dz  |\psi(z)|^2 e^{i q z} 
\label{eq:formfactor}
\end{equation}
where $\psi(z)$ is the exciton center-of-mass envelope wave function and where $\omega=v|q|$. 
Note that in this Ohmic model one has $F_0(0)=D$ regardless of the exciton wave function, which leads to the merging of the ZPL into the phonon wings.
In the case of a Gaussian exciton envelope,  $\psi(z)=\frac{1}{\pi^{1/4}\sigma^{1/2}} \exp(-\frac{z^2}{2\sigma^2})$, where $\sigma$ is the exciton localization length, the form factor reads :
\begin{equation}
 F_0 (\omega)=D\exp[-(\omega\sigma/2v)^2]
\label{eq:formfactorgauss}
 \end{equation}
This implies a high energy cut-off of the exciton-acoustic phonon coupling at $\hbar \omega_\sigma \approx 2 \hbar v /\sigma$ that simply results from momentum conservation within the width of the exciton envelope in the $k$-space. 

The emission spectra can be computed from the coupling matrix element by summing up the contributions of all phonons modes weighted by their occupation numbers \cite{Krummheuer2002,Galland2008}. 
Conversely, the form factor $F(\omega)$ can be traced back from the experimental spectra with the procedure described in \cite{supmat}. 
The form factor extracted from the broad and asymmetric profile is presented in Fig.~\ref{fig1}~(e). This form factor fits well to a Gaussian such as $F_0$, as expected in the Ohmic model \cite{Galland2008}.

Even if the experimental spectra show a large variety of profiles (Fig. \ref{fig1} (a-d)), the extracted form factors share common features. Importantly, we note that they do not depend on the temperature (as expected from the model). In addition, all the form factors display a Gaussian baseline, from which we can extract a deformation potential $D$ between 12 and 15 eV, and an exciton localization length $\sigma$ of around 3 nm, similar to the ones found in \cite{Galland2008}. 
The main deviation from the Ohmic coupling occurs for low phonon energies, typically below 3 meV. For instance the narrow spectrum of Fig. 1 (b) corresponds to a complete suppression of the coupling ($F(\omega)=0$) for low energy phonons (Fig. 1 (f)). 
Again, we stress that this reduced coupling cannot be accounted for by a modification of the exciton wave function in the Ohmic model, since the form factor goes to $D$ for vanishing $\omega$ regardless of the envelope shape (Eq. \ref{eq:formfactor}). In terms of emission spectrum, this means that changing the exciton envelope function (shape or localization length) cannot account simultaneously for the observed narrow ZPL (requiring a delocalized exciton) and a broader pedestal (requiring a tightly localized exciton). 
Therefore, the low-energy modification of the form factor must arise from the phonon modes themselves (as discussed at the end of the paper).

The key to the understanding of the strong variations of the PL line shapes actually lies in the low-energy exciton-phonon coupling. When $F$ takes sizable values for vanishing phonon energies, the ZPL is strongly suppressed to the profit of the phonon wing. In contrast, if the form factor is reduced for low phonon energies ($\omega<\omega_c$) a sharp ZPL is restored (Fig.~\ref{fig1}~(b)). 
Practically, this effect is observable if the cut-off phonon energy $\hbar \omega_c$ is greater than the natural line-width of the ZPL : $\omega_c > 2/T_2$, where $T_2$ is the ZPL dephasing time. Experimentally, 
we found a width of $\approx 0.65$~meV for the ZPL and $\hbar\omega_c \approx 1.2$~meV in Fig. 1 (a), which explains why the phonon sideband and the ZPL are not completely split off. 

\begin{figure}
\includegraphics[width=8.6cm]{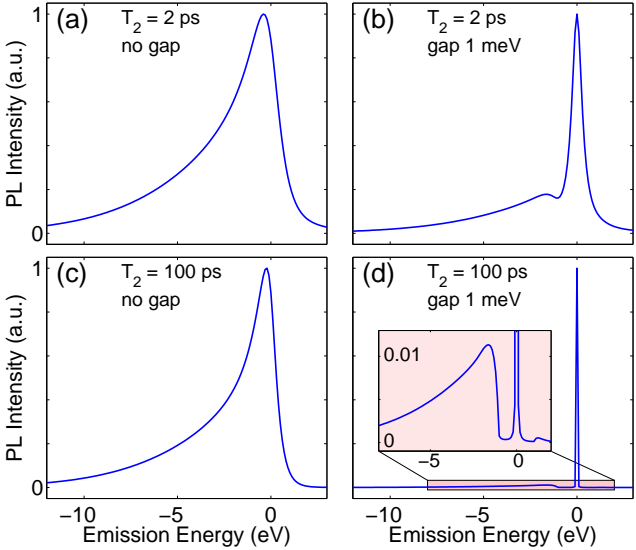}
\caption{Calculated spectrum for the Ohmic coupling (panels (a) and (c)) and for a 1 meV gap in the coupling (panels (b) and (d)). Spectrum are calculated assuming a dephasing time $T_2$ of 2 ps (panels (a) and (b)), and for $T_2=$ 100 ps (panels (c) and (d)). All the other parameters are identical for the four plots ($T=5$~K, $D=13$~eV and $\sigma=3$~nm). \label{fig:simulT2}}
\end{figure}

More generally, the effect of the dephasing time on the spectral line shape is presented in Fig.~\ref{fig:simulT2}. In the case of the Ohmic coupling, the line shape is roughly insensitive to the value of the electronic dephasing time. This is a peculiarity of the 1D geometry, where the ZPL is completely merged into the phonon wings. In the case of a low-energy gap in the form factor (in Fig.~\ref{fig:simulT2} we used a 1~meV gap), a sharp ZPL is restored. 
The amplitude of the ZPL is directly proportional to the dephasing time $T_2$ since the area of the line must be constant, reflecting the oscillator strength of the transition. 
In contrast the phonon wings hardly depend on $T_2$. Longer dephasing times such as those observed in suspended nanotubes therefore lead to an apparent magnification of the ZPL. 
For example in the Fig.~1 of \cite{Hofmann2013}, a sharp ZPL and a weak phonon sideband are observable \footnote{Ultra-narrow PL spectra of suspended nanotubes at 4 and 77 K provided by A. Hoegele in a private communication could be quantitatively fitted to our model with an acoustic barrier width of 80 nm and height of 1 meV}. 
The ZPL / sideband amplitude ratio is of the order of 100 for a ZPL width of 40 $\mu$eV. Consistently, the spectrum displayed in Fig. \ref{fig1} (b) shows a ratio of the order of 5, for a ZPL width of 600 $\mu$eV. 

We note that some suspended NT spectra have no apparent phonon wings. We believe that this is due to experimental limitations. In fact, reaching a given signal to noise ratio requires much shorter integration times for sharp ZPLs, which may loose the weak sidebands in the noise background. \footnote{In some other cases where no anti-bunching signature is observed \cite{Sarpkaya2013}, the exciton could be more delocalized, which would also suppress the wings.}

Let us now examine possible origins of the reduction of the coupling to the low-energy phonon modes. 
Equivalently, this low-energy cut-off can be seen as a reduction of the low energy phonon density of states probed by the exciton. 
Generally speaking, such local modifications of the phonon modes can be accounted for by introducing acoustic barriers, with height $\hbar\omega_c \simeq$~2~meV \cite{supmat}. 
Actually, since 1D structures are much more sensitive to perturbations than their higher dimension counterparts, several ordinary effects may cause important phonon modifications.  In the case of suspended SWNTs, amorphous carbon deposited by places or structural defects can break the 1D translational symmetry, 
leading to low energy phonon modifications \footnote{We note that self trapping of the exciton with acoustic phonon can also lead to a modification of the phonon spectrum in the meV range \cite{Shaw1981}}.
In the case of embedded SWNTs, the non-covalent contact to the substrate or matrix is known to occur on nanometer scale zones \cite{Crochet2012} and leads to the opening of few-meV-phonon gaps \cite{Savin2009} due to a local hardening of the nanotube lattice. 

The different types of PL profiles then arise from different geometries of the barrier. 
We make use of the general expression for the form factor $F$ :
\begin{equation}
F(\omega_i)=D \  \sqrt{L}\  v/\omega_i\  \int dz  |\psi(z)|^2 W_i'(z)
\end{equation}
where $W_i(z)$ are the orthonormal acoustic modes, solutions of :
\begin{equation}
-\omega_i^2 W_i(z)=v^2 W_i''(z) - \omega_b^2(z) W_i(z)
\end{equation}
For the sake of simplicity, we assume here that the acoustic barrier amplitude $\hbar\omega_b(z)$ is either 0 or $\hbar\omega_c$ along the nanotube (fig.~\ref{fig1}(i-l) . 
Without barrier, free propagating modes are obtained and $F=F_0$ (Fig.~\ref{fig1} (a) and (e)).
If the barrier width is much larger than the exciton localization length, a true gap is found in the form factor (Fig.~\ref{fig1} (b) and (f)). 
For an intermediate barrier width, typically a few tens of nanometers, the coupling is only partially canceled, owing to the finite attenuation length of the evanescent low energy phonon modes inside the barrier (Fig.~\ref{fig1} (c) and (g)) \cite{supmat}.

An additional case of interest corresponds to the situation where the exciton is located between two acoustic barriers separated by a few tens of nanometers. Our model predicts that in this case side-peaks should appear in the phonon wing because the exciton is coupled to a resonant vibration mode between the barriers. We actually found experimental evidences for such resonant coupling (Fig.~\ref{fig1}~(d) and (h)). 
The observed splitting (2.5 meV) corresponds to a distance between the barriers of the order of 20 nm, consistent with the scale of variations of the environment of the nanotube. This original observation corresponds to the interesting situation of disorder-induced confinement of vibrations at the nanoscale.

In conclusion, the low temperature PL profiles of carbon nanotubes (ranging from sharp sub-meV lines to broad asymmetrical lines) are well accounted for within a single model describing the coupling of a localized exciton and a one-dimensional phonon bath. The key to the diversity of the profiles lies in tiny modifications of the low-energy phonons.
The observation of a narrow lineshape indicates the presence of a gap in the exciton-phonon coupling.
Furthermore, we pointed out the key role of the ZPL dephasing rate in the overall profile of the PL, explaining the apparent discrepancy between suspended and embedded nanotubes. This
reduction of the coupling to acoustic phonons is a first important step towards the control of the emission line-width of matrix embedded carbon nanotubes. 
Finally, we showed that phonons may be confined by disorder with specific fingerprints in the PL spectrum. 

\subsection*{Acknowledgement}
This work was supported by the GDR-I GNT, the grant \textit{"C'Nano IdF TENAPO"}. CV and GC are members of ``Institut Universitaire de France''.\\



\bibliography{phonon_wing_paper}

\newpage
\part*{Supplemental Material for : ``Unifying the low-temperature photoluminescence spectra of carbon nanotubes: the role of acoustic phonon confinement''.}

\setcounter{figure}{0}
\renewcommand{\thefigure}{S-\arabic{figure}}


\section{Single emitter}

The sample was characterized by Atomic Force Microscopy (AFM). We observed surfactant aggregates along with a few well individualized tubes. 
Their length ranges from 100 nm to 1 $\mu$m.
Since the density of SWNTs can strongly vary from one location to another at the surface of the sample, optical studies were performed only in zones with a rather low density of emitting tubes ($\simeq 10^{-1} \mu$m$^{-2}$).

At cryogenic temperatures, the PL signals exhibit blinking, spectral jumps and spectral diffusion which are typical of single emitters. 
Extensive optical studies were performed on the most stable ones.
In the following we present observations made on the SWNTs described in Fig. 1 (c) and Fig. 1 (d) of the paper. 
Similar behaviors were observed for several dozens of SWNTs. All the spectra were recorded at 10K.

\begin{figure}[!ht]
\includegraphics[height=6.5cm]{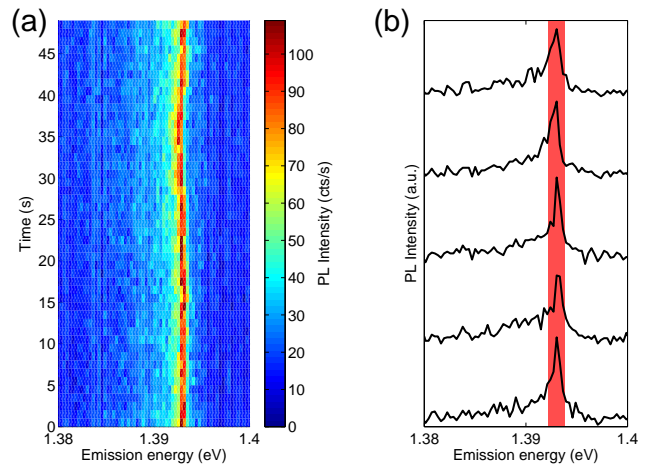}
\caption{(a) Time trace of the PL spectrum of the nanotube presented in Fig. 1 (c), recorded every second. 
(b)  PL spectra recorded every 10~s with a 1~s exposure time. The color stripe shows the integration energy range (1.5 meV) used for Fig. S-3. 
\label{fig:interm-time}.}
\end{figure}

\begin{figure}[!ht]
\includegraphics[height=7cm]{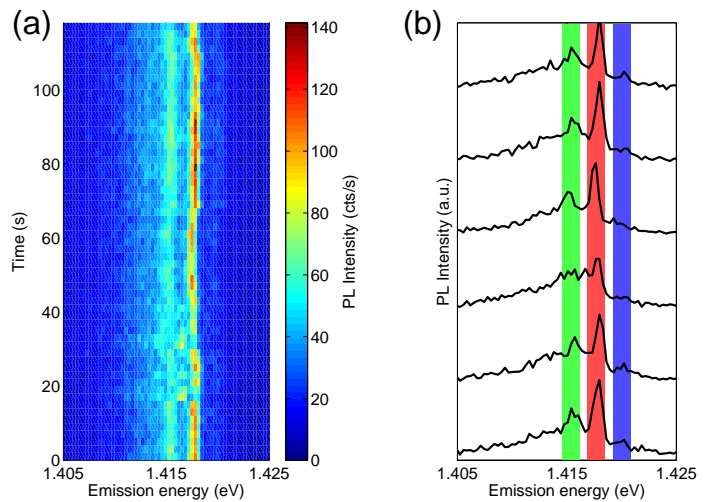}
\caption{(a) Time trace of the PL spectrum of the nanotube presented in Fig. 1 (d), recorded every 2 seconds. 
(b)  PL spectra recorded every 20~s with a 2~s exposure time. The color stripes show the integration energy range (1.5 meV) used for Fig. S-4-5. 
\label{fig:multi-time}.}
\end{figure}

Long time traces with short exposure time steps show a rather stable emission energy, with a spectral diffusion smaller than the linewidth, and a stable emission intensity (Fig.~\ref{fig:interm-time} and Fig.~\ref{fig:multi-time}). In the particular case of the multi-peak spectrum, we observe that the magnitude of the spectral diffusion (2 pixels = 0.6 meV) remains much weaker than the typical splitting between the components of the spectrum (of the order of 8 pixels) and that this splitting is constant up to 1 pixel (0.3meV) whatever the absolute position of the lines.

\begin{figure}[!ht]
\includegraphics[height=5cm]{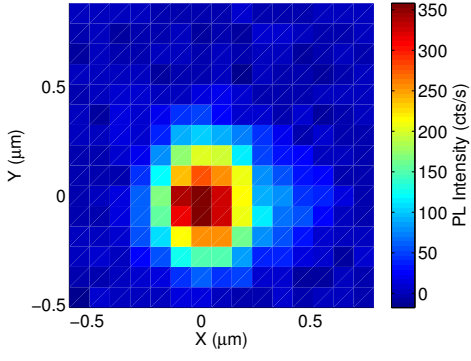}
\caption{Confocal spatial PL map of the nanotube in Fig. 1 (c). For each point of the map, a PL spectrum was recorded with a 1 s exposure time and the integrated intensity over the energy range given in Fig.~\ref{fig:interm-time}(b) is then reported according to the color scale. Spatial steps are 100 nm long 
\label{fig:interm-map}.}
\end{figure}

\begin{figure}[!ht]
\includegraphics[height=15cm]{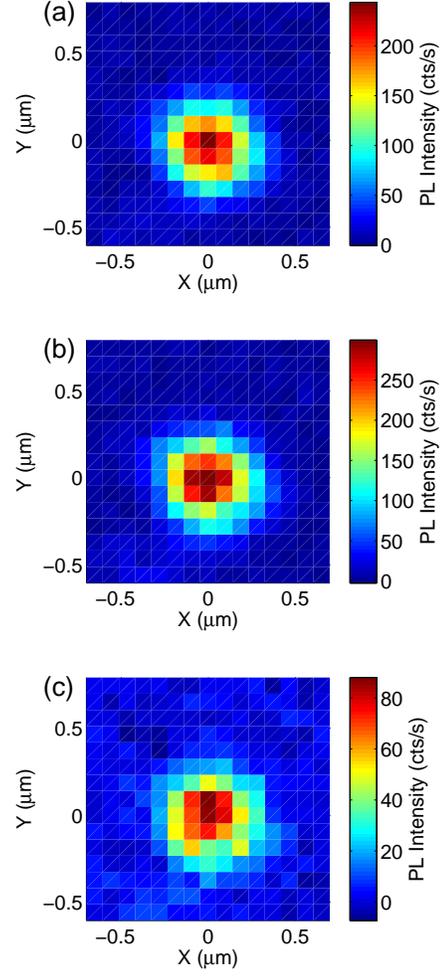}
\caption{Confocal PL maps of the 3 different peaks observed in the spectrum of the nanotube in Fig. 1 (d). For each point of the map, a single PL spectrum was recorded with a 2 s exposure time and the integrated intensity over each energy range given in Fig.~\ref{fig:multi-time}(b) is then reported according to the color scale. (a) (b) and (c) correspond respectively to the peak at 1.416 eV, 1.418 eV and 1.42 eV. Spatial steps are 100 nm long 
\label{fig:multi-map}.}
\end{figure}

In addition, spatial PL maps show clear localized emission spots (Fig.~\ref{fig:interm-map} and Fig.~\ref{fig:multi-map}). 
A 2-dimensional Gaussian fit gives a full width at half-maximum of 410 $\pm$ 30 nm, close to the diffraction limited resolution of the setup.
However, one cannot discriminate between a nanometer scale localized exciton in a long tube and a more delocalized exciton in a non resolved very short nanotube.

\begin{figure}[!ht]
\includegraphics[height=6cm]{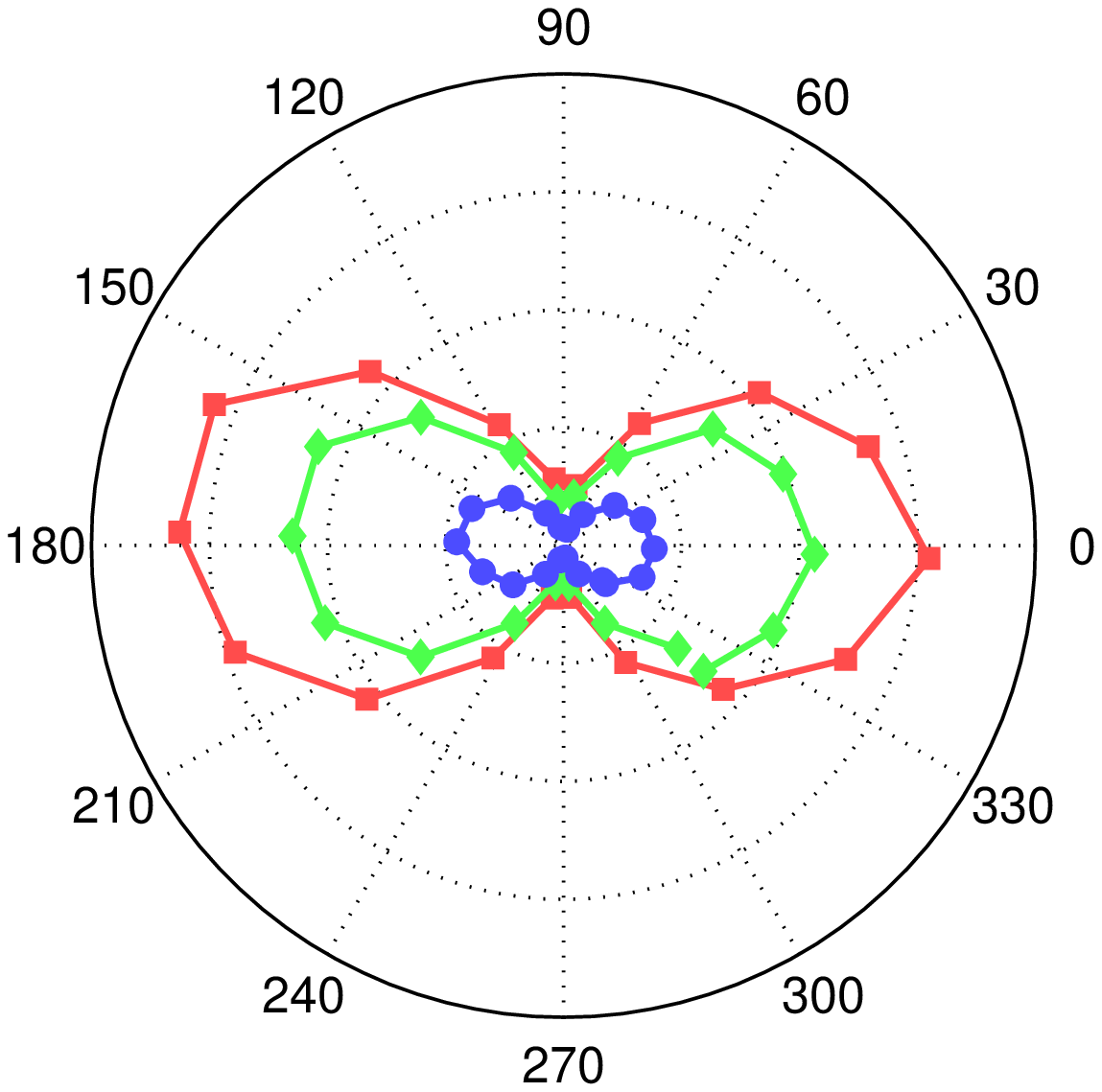}
\caption{Excitation polarization diagram of the integrated PL of the nanotube of Fig. 1 (d). The 3 different peaks are represented according to their respective color attributed in Fig.~\ref{fig:multi-time}(b)
\label{fig:polar}.}
\end{figure}

Moreover, we always observed strongly anisotropic excitation and emission polarization diagrams (Fig.~\ref{fig:polar}), which is typical of nanotubes for which the oscillator strength is much larger for a polarization along the tube axis. 

All these observations support the fact that the spectra arise from single nanotubes rather than from several emitters. 

Special care has been taken to come up to this conclusion in the case of the multi-peak spectrum.
Indeed, spatial maps of the 3 different peaks show a perfect overlap within the Gaussian fit error bars of 30 nm (Fig.~\ref{fig:multi-map}). 
In addition, we can note the correlated polarization behaviors (Fig.~\ref{fig:polar}) and correlated emission intensities over time (Fig.~\ref{fig:multi-time}(a)).

Importantly, we note that the characteristic asymmetrical and multi-peak profiles reported in the paper clearly show up for any integration time. Together with the singular temperature dependence of the spectra, we can safely rule out that these feature arise from spectral diffusion.


\section{Branching ratio between ZPL and phonon wing}

The Hamiltonian of a two-level excitonic system coupled to an acoustic phonon mode reads:
\begin{equation}
H=\hbar \Omega |e\rangle \langle e|  +\hbar \omega_i b_i^\dagger b_i  + \hbar G_i |e\rangle \langle e| \otimes (b_i +b_i^{\dagger} )
\end{equation}
where $|g\rangle$ and $|e\rangle$ are the ground and excited electronic state, $b_i$ (resp. $b_i^\dagger$)  is the annihilation (resp. creation) acoustic phonon operator, $\hbar\Omega$ is the energy difference between the ground and excited electronic state and $\hbar\omega_i$ is the phonon energy. The  bulk electron - acoustic phonon coupling $G_i$ is given by: $G_i=D \sqrt{\frac{\omega_i}{2 \rho V v^2 \hbar}}$, where $D$ is the deformation potential, $\rho$  the mass density, $V$ a normalisation volume and $v$ the sound speed (for an unidimensional system, $\rho$ is the linear mass density and $V$ should be replace by $L$ the length of the system). 
H can be diagonalized into:
\begin{equation}
H=\hbar \omega_i |g\rangle \langle g|\otimes b_i^\dagger b_i \ +\  |e\rangle \langle e| \otimes (\hbar \Omega + \hbar\omega_i D_\xi^\dagger b_i^\dagger b_i D_\xi )  - \hbar\omega_i \xi^2
\end{equation}
where $\xi=G_{i}/\omega_i$ and $D_\xi$ is the displacement operator given by $D_\xi=\exp(\xi b^\dagger_i-\xi b_i)$. 
The eigenstates of this hamiltonian are given by $|g,n \rangle$ and $D_\xi^\dagger |e,n \rangle$. 
\begin{figure}[!ht]
\includegraphics[height=5cm]{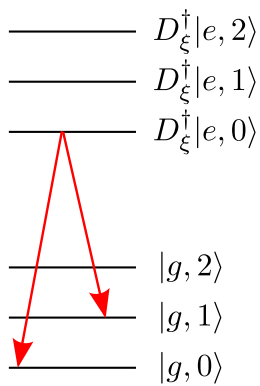}
\caption{Energy diagram of a two level system coupled to an acoustic phonon mode\label{fig:branching}.}
\end{figure}
The probability of optical transition is proportional to $\left| \langle g,n' | \ d\  | D_\xi^\dagger|e,n\rangle \right|^2$, where $d$ is the dipolar operator acting only on the electronic part of the wavefunction. 
The branching ratio ($br$) between the emission of a photon-phonon pair (\textit{i.e.} in the wings) and the emission of a photon alone (\textit{i.e}. in the ZPL) (cf Fig. \ref{fig:branching}) is given by:
\begin{equation}
br_i=\left|\frac{\langle 1 |D_\xi^\dagger | 0 \rangle}{\langle 0 |D_\xi^\dagger | 0 \rangle}\right|^2 = \frac{\xi^2 \exp(|\xi|^2)}{ \exp(|\xi|^2)}=\frac{G_i^2}{\omega_i^2}
\end{equation}

When considering the total coupling to all phonon modes, the previous result is multiplied by the density of acoustic phonon states, given by:  $DOS \propto \omega^{Dim-1}$, where $Dim$ is the dimensionality  of the phonon bath.
From the dependency  of the coupling $G_i$ with $\omega_i$, one obtains that the branching ratio goes as :
\begin{equation}
br\propto \omega^{Dim-2}
\end{equation}

This ratio becomes diverging at low energy for a one-dimensional phonon bath, leading to a complete merging of the ZPL into the phonon wings.


\section{Temperature dependence of the phonon wings}

\begin{figure}[!ht]
\includegraphics[height=10cm]{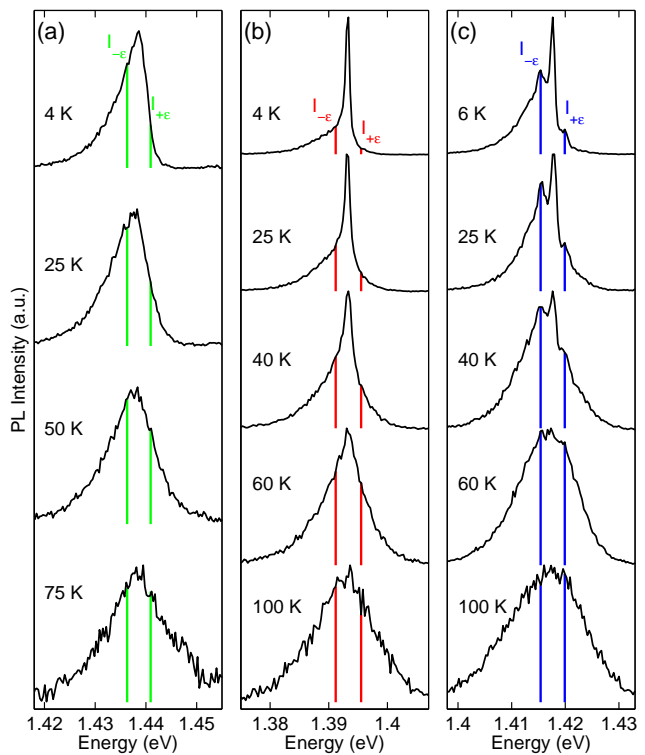}
\caption{Evolution of 3 different spectra presented in the main article as a function of the temperature. 
Spectra are normalized and shifted vertically for clarity.
The PL intensities $I_{-\epsilon}$ and $I_{+\epsilon}$ are represented by a vertical bar for an energy shift $\epsilon$ of  $2.3$~meV. 
\label{fig:temp1}}
\end{figure}

\begin{figure}[!ht]
\includegraphics[height=6cm]{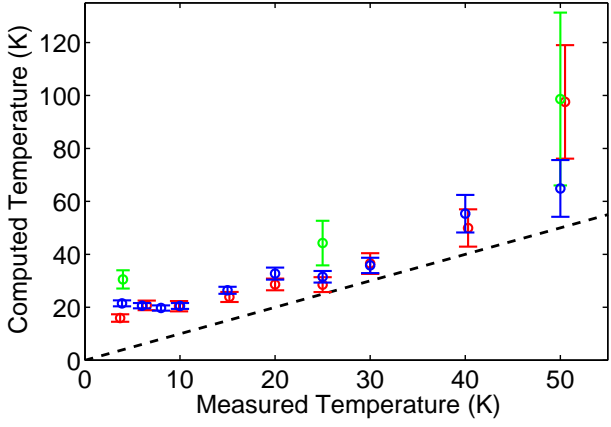}
\caption{Evolution as a function of the experimental temperature of the effective temperature evaluated with a Boltzmann factor : $T_{eff}=\epsilon / k_B \, \ln(I_{-\epsilon}/I_{+\epsilon})$. The color code is identical to the one of Fig. \ref{fig:temp1}.
\label{fig:temp2}.}
\end{figure}

For each spectrum, we evaluate the amplitude ratio between the blue and the red wings.
We introduce the ratio $I_{+\epsilon}/I_{-\epsilon}$ where $I_{+\epsilon}$ (resp. $I_{-\epsilon}$) is the PL intensity at an energy $E = E_{max} + \epsilon$ (resp. $E=E_{max} - \epsilon$), $E_{max}$ being the energy at maximum PL intensity (Fig. \ref{fig:temp1}).
In a simple single phonon coupling interpretation, the blue wing involves the creation of an acoustic phonon proportional to $n_{\epsilon}$ the occupation number of a phonon mode with energy $\epsilon$.
Conversely, the red wing involves the annihilation of a phonon and is proportional to $n_{\epsilon}+1$.
We obtain $I_{+\epsilon}/I_{-\epsilon} = n_{\epsilon}/(n_{\epsilon}+1) = \text{exp}(-\epsilon/k_BT)$.
An effective temperature $T_{eff}$ can be deduced from this relation.
We note that $T_{eff}$ is accurately evaluated only for temperatures $T$ lower than a few $\epsilon / k_B (\sim 30$~K) since a $I_{+\epsilon}/I_{-\epsilon}$ ratio close to 1 (within the PL noise) gives diverging uncertainties for $T_{eff}$.
We find a good agreement between $T_{eff}$ and the experimentally measured temperatures (Fig. \ref{fig:temp2}).
The slightly higher values of $T_{eff}$, especially at very low temperature ($T\leq10$~K) may come from local heating due to the excitation laser beam. We note that the temperature dependence is better reproduced by the full model (as shown in the main text) because it includes the ZPL dephasing time, which tends to smooth the spectra and also because it includes the contribution of multi-phonons emission (or absorption), which generates deviations from the simple Boltzmann factor.

\section{Form factor: numerical estimation}
Following the methods presented in \cite{Krummheuer2002,Galland2008}, the spectrum can be computed from the form factor. The relationship between the spectrum and the form factor is a mathematically reversible operation, and thus it ensures that a given spectrum corresponds to an unique form factor. We note that the noise on the experimental spectrum  leads to numerical instabilities in the inversion process. Therefore, we choosed to use a random optimization procedure to extract the form factor. We initialized the procedure with a flat form factor. Then, by means of an iterative procedure, we searched for random modifications that make the computed spectrum closer to the experimental one. This procedure gives reproducible form factors, at the cost of some high frequency noise.


\section{Acoustic barriers}

The electron - acoustic phonon coupling Hamiltonian is given by :
\begin{equation}
H_{e-AP}=D \,\mathrm{div}(u)
\end{equation}
where the displacement $u$ reads :
\begin{equation}
u=\sum_i \sqrt{\frac{\hbar}{2\rho\omega_i}} (b_i+b_i^{\dagger}) W_i(z)
\end{equation}
The $W_i(z)$ form an orthonormal basis of the acoustic modes.
For an exciton, the coupling Hamiltonian $H_{X-AP}$ can be simplified in :
\begin{equation}
\langle \Psi_X | D \,\text{div}(u) | \Psi_X \rangle = \sum_i  \hbar g_{\omega_{i}} ( b_i+b_i^{\dagger} ) 
\end{equation}
with the matix element $g_{\omega_{i}}$ given in the main text.
\begin{equation}
\hbar g_{\omega_{i}}= \sqrt{\frac{\hbar \omega_i}{2\rho L v^2}} F(\omega_i)
\end{equation}
where $F(\omega_i)$ reads :
\begin{equation}
F(\omega_i)=D \  \sqrt{L}\  v/\omega_i\  \left\langle \Psi_X \left| \frac{dW_i(z)}{dz} \right| \Psi_X \right\rangle
\end{equation}

To find the acoustic mode $W_i$ one has to solve the following equation :
\begin{equation}
-\omega_{i}^2 W_i(z)=v^2 \frac{d^2 W_i(z)}{dz^2} -
	\begin{cases}
		0 &\text{, no barrier} \\
		\omega_c^2 W_i(z) & \text{, barrier}\\
	 \end{cases}
\end{equation}

The free propagating mode corresponds to the upper line, whereas the lower one describes the acoustic barrier due to the local coupling to the environment (for instance a local increase of the spring constant).
The discretization of the eigen frequencies $\omega_{i}$ and eigen modes $W_{i}$ come from the finite length of the tube $L=1$~$\mu$m in the calculation.
For the sake of simplicity, we assume that the acoustic barrier amplitude is either 0 or $\hbar\omega_c$ along the length of the nanotube.

Once the acoustic modes $W_i$ are normalized, the form factor $F$ can be computed.
Simplified analytical expressions can be obtained for simple symmetrical configurations with the exciton at the centre (position $z=0$ in Fig.~1~(i-l)).
The free propagating case (no acoustic barrier $F=F_0$) and the case of an infinitely long acoustic barrier ($F=F_\infty$) are represented in  Fig.~1~(i) and (j) respectively.
Approximated expressions can be obtained when the barrier length $a$ is finite but much larger than the exciton size $\sigma$ ($L>>a>>\sigma$, $F_a$, Fig. 1 (k)). Similarly, an approximated analytical expression can be derived for two identical barriers (of width $a$) spaced by a distance $l$ ($F_{l,a}$, Fig.~1~(l)). 
\begin{eqnarray}
F_0(\omega)      &=& D \:\text{exp}\left[-  \frac{\sigma^2}{4v^2} \, \omega^2  \right]\\
F_\infty (\omega)  &=& F_0(\tilde{\omega}) \frac{Re[\tilde{\omega}]}{\omega}  \\
F_a(\omega)     &\approx &  F_0(\tilde{\omega})\left(1+ \frac{\omega_c^2 sin^2(a \tilde{\omega}/2v)}{\tilde{\omega}^2} \right)^{-1/2}
\end{eqnarray}

\begin{multline}
F_{l,a}(\omega)  \approx F_0(\omega) \left( \left| \text{Cos}\alpha \, \text{Sin}\beta + \frac{\omega}{\tilde{\omega }} \,\text{Sin}\alpha \,\text{Cos}\beta \right|^2 \right. \\
+  \left. \left| \text{Cos}\alpha \,\text{Cos}\beta - \frac{\tilde{\omega }}{\omega } \,\text{Sin}\alpha \,\text{Sin}\beta \right|^2 \right)^{-1/2}
\end{multline}

Where $\tilde{\omega}$ is the complex number $\tilde{\omega}=\sqrt{\omega^2-\omega_c^2}$, and $\alpha$ and $\beta$ are given by $\alpha=a\tilde{\omega}/v$ and $\beta=l\omega/2v$. 

For each case, a fit to the form factor extrated from experimental spectra presented in Fig. 1 in the main text is performed.
The best fits were obtained with the set of parameters listed in Table~\ref{tbl:param}.
The deformation potential $D$ and exciton enveloppe size $\sigma$ are similar to the one evaluated with the Gaussian baseline.
The barrier height $\hbar\omega_c$ is found to be between 1 and 2~meV.
The spatial structures (barrier widths and distance between barriers) are of the order of 10 to 35 nanometers.
These fits to the form factor were further used to simulate the emission spectra with an excellent accuracy, as shown in Fig. 1 (a)-(d) of the main text.
\begin{table}[!ht]
\setlength{\tabcolsep}{10pt}
\begin{center} 
\begin{tabular}{c c c c c c|}

 \hline \hline
 
 Figure 1 & (e) & (f) & (g) & (h) \\
 
 \hline
 
 $D$ [eV] &  14 & 12 & 12 & 15 \\
 $\sigma$ [nm] &  2.4 & 2.4 & 2.6 & 3.0 \\
 $\hbar\omega_c$ [meV] &  - & 1.1 & 1.3 & 1.9 \\ 
 $a$ [nm] & - & - & 34 & 8 \\
 $l$ [nm] & - & - & - & 20 \\
 
 \hline \hline
 \end{tabular}
\end{center}
\caption{Best fit parameters for the data displayed in Fig. 1. The acoustic barrier is defined with its height $\hbar\omega_c$ and width $a$. 
In the case of the multipeak spectrum (Fig. 1~(d)), the gap width between the two identical barriers is $l$.} 
\label{tbl:param}
\end{table}

Finally, we can deduce from the statistical occurrence of the different cases presented in Figure~1, that there is no preferential localization sites of the exciton with respect of the mechanical contact zones. Among a few tens of specimens, we found that most spectra fall in the category of Figure~1 (c), corresponding to an intermediate coupling scheme between the limit cases (a) and (b). The observation of confined phonon modes such as in Figure~1 (d) represents less than 10\% of the cases.

\bibliography{phonon_wing_paper}

\begin{thebibliography}{22}%
\makeatletter
\providecommand \@ifxundefined [1]{%
 \@ifx{#1\undefined}
}%
\providecommand \@ifnum [1]{%
 \ifnum #1\expandafter \@firstoftwo
 \else \expandafter \@secondoftwo
 \fi
}%
\providecommand \@ifx [1]{%
 \ifx #1\expandafter \@firstoftwo
 \else \expandafter \@secondoftwo
 \fi
}%
\providecommand \natexlab [1]{#1}%
\providecommand \enquote  [1]{``#1''}%
\providecommand \bibnamefont  [1]{#1}%
\providecommand \bibfnamefont [1]{#1}%
\providecommand \citenamefont [1]{#1}%
\providecommand \href@noop [0]{\@secondoftwo}%
\providecommand \href [0]{\begingroup \@sanitize@url \@href}%
\providecommand \@href[1]{\@@startlink{#1}\@@href}%
\providecommand \@@href[1]{\endgroup#1\@@endlink}%
\providecommand \@sanitize@url [0]{\catcode `\\12\catcode `\$12\catcode
  `\&12\catcode `\#12\catcode `\^12\catcode `\_12\catcode `\%12\relax}%
\providecommand \@@startlink[1]{}%
\providecommand \@@endlink[0]{}%
\providecommand \url  [0]{\begingroup\@sanitize@url \@url }%
\providecommand \@url [1]{\endgroup\@href {#1}{\urlprefix }}%
\providecommand \urlprefix  [0]{URL }%
\providecommand \Eprint [0]{\href }%
\providecommand \doibase [0]{http://dx.doi.org/}%
\providecommand \selectlanguage [0]{\@gobble}%
\providecommand \bibinfo  [0]{\@secondoftwo}%
\providecommand \bibfield  [0]{\@secondoftwo}%
\providecommand \translation [1]{[#1]}%
\providecommand \BibitemOpen [0]{}%
\providecommand \bibitemStop [0]{}%
\providecommand \bibitemNoStop [0]{.\EOS\space}%
\providecommand \EOS [0]{\spacefactor3000\relax}%
\providecommand \BibitemShut  [1]{\csname bibitem#1\endcsname}%
\let\auto@bib@innerbib\@empty
\bibitem [{\citenamefont {Wang}\ \emph {et~al.}(2005)\citenamefont {Wang},
  \citenamefont {Dukovic}, \citenamefont {Brus},\ and\ \citenamefont
  {Heinz}}]{Wang2005}%
  \BibitemOpen
  \bibfield  {author} {\bibinfo {author} {\bibfnamefont {F.}~\bibnamefont
  {Wang}}, \bibinfo {author} {\bibfnamefont {G.}~\bibnamefont {Dukovic}},
  \bibinfo {author} {\bibfnamefont {L.~E.}\ \bibnamefont {Brus}}, \ and\
  \bibinfo {author} {\bibfnamefont {T.~F.}\ \bibnamefont {Heinz}},\ }\href
  {\doibase 10.1126/science.1110265} {\bibfield  {journal} {\bibinfo  {journal}
  {Science}\ }\textbf {\bibinfo {volume} {308}},\ \bibinfo {pages} {838}
  (\bibinfo {year} {2005})}\BibitemShut {NoStop}%
\bibitem [{\citenamefont {Matsunaga}\ \emph {et~al.}(2011)\citenamefont
  {Matsunaga}, \citenamefont {Matsuda},\ and\ \citenamefont
  {Kanemitsu}}]{Matsunaga2011}%
  \BibitemOpen
  \bibfield  {author} {\bibinfo {author} {\bibfnamefont {R.}~\bibnamefont
  {Matsunaga}}, \bibinfo {author} {\bibfnamefont {K.}~\bibnamefont {Matsuda}},
  \ and\ \bibinfo {author} {\bibfnamefont {Y.}~\bibnamefont {Kanemitsu}},\
  }\href {\doibase 10.1103/PhysRevLett.106.037404} {\bibfield  {journal}
  {\bibinfo  {journal} {Phys. Rev. Lett.}\ }\textbf {\bibinfo {volume} {106}},\
  \bibinfo {pages} {037404} (\bibinfo {year} {2011})}\BibitemShut {NoStop}%
\bibitem [{\citenamefont {Colombier}\ \emph {et~al.}(2012)\citenamefont
  {Colombier}, \citenamefont {Selles}, \citenamefont {Rousseau}, \citenamefont
  {Lauret}, \citenamefont {Vialla}, \citenamefont {Voisin},\ and\ \citenamefont
  {Cassabois}}]{Colombier2012}%
  \BibitemOpen
  \bibfield  {author} {\bibinfo {author} {\bibfnamefont {L.}~\bibnamefont
  {Colombier}}, \bibinfo {author} {\bibfnamefont {J.}~\bibnamefont {Selles}},
  \bibinfo {author} {\bibfnamefont {E.}~\bibnamefont {Rousseau}}, \bibinfo
  {author} {\bibfnamefont {J.~S.}\ \bibnamefont {Lauret}}, \bibinfo {author}
  {\bibfnamefont {F.}~\bibnamefont {Vialla}}, \bibinfo {author} {\bibfnamefont
  {C.}~\bibnamefont {Voisin}}, \ and\ \bibinfo {author} {\bibfnamefont
  {G.}~\bibnamefont {Cassabois}},\ }\href
  {http://link.aps.org/doi/10.1103/PhysRevLett.109.197402} {\bibfield
  {journal} {\bibinfo  {journal} {Phys. Rev. Lett.}\ }\textbf {\bibinfo
  {volume} {109}},\ \bibinfo {pages} {197402} (\bibinfo {year}
  {2012})}\BibitemShut {NoStop}%
\bibitem [{\citenamefont {H\"ogele}\ \emph {et~al.}(2008)\citenamefont
  {H\"ogele}, \citenamefont {Galland}, \citenamefont {Winger},\ and\
  \citenamefont {Imamoglu}}]{Hogele2008}%
  \BibitemOpen
  \bibfield  {author} {\bibinfo {author} {\bibfnamefont {A.}~\bibnamefont
  {H\"ogele}}, \bibinfo {author} {\bibfnamefont {C.}~\bibnamefont {Galland}},
  \bibinfo {author} {\bibfnamefont {M.}~\bibnamefont {Winger}}, \ and\ \bibinfo
  {author} {\bibfnamefont {A.}~\bibnamefont {Imamoglu}},\ }\href {\doibase
  10.1103/PhysRevLett.100.217401} {\bibfield  {journal} {\bibinfo  {journal}
  {Phys. Rev. Lett.}\ }\textbf {\bibinfo {volume} {100}},\ \bibinfo {eid}
  {217401} (\bibinfo {year} {2008})}\BibitemShut {NoStop}%
\bibitem [{\citenamefont {Hofmann}\ \emph {et~al.}(2013)\citenamefont
  {Hofmann}, \citenamefont {Gluckert}, \citenamefont {Noe}, \citenamefont
  {Bourjau}, \citenamefont {Dehmel},\ and\ \citenamefont
  {H\"ogele}}]{Hofmann2013}%
  \BibitemOpen
  \bibfield  {author} {\bibinfo {author} {\bibfnamefont {M.~S.}\ \bibnamefont
  {Hofmann}}, \bibinfo {author} {\bibfnamefont {J.~T.}\ \bibnamefont
  {Gluckert}}, \bibinfo {author} {\bibfnamefont {J.}~\bibnamefont {Noe}},
  \bibinfo {author} {\bibfnamefont {C.}~\bibnamefont {Bourjau}}, \bibinfo
  {author} {\bibfnamefont {R.}~\bibnamefont {Dehmel}}, \ and\ \bibinfo {author}
  {\bibfnamefont {A.}~\bibnamefont {H\"ogele}},\ }\href
  {http://dx.doi.org/10.1038/nnano.2013.119} {\bibfield  {journal} {\bibinfo
  {journal} {Nat Nano}\ }\textbf {\bibinfo {volume} {8}},\ \bibinfo {pages}
  {502} (\bibinfo {year} {2013})}\BibitemShut {NoStop}%
\bibitem [{\citenamefont {Sarpkaya}\ \emph {et~al.}(2013)\citenamefont
  {Sarpkaya}, \citenamefont {Zhang}, \citenamefont {Walden-Newman},
  \citenamefont {Wang}, \citenamefont {Hone}, \citenamefont {Wong},\ and\
  \citenamefont {Strauf}}]{Sarpkaya2013}%
  \BibitemOpen
  \bibfield  {author} {\bibinfo {author} {\bibfnamefont {I.}~\bibnamefont
  {Sarpkaya}}, \bibinfo {author} {\bibfnamefont {Z.}~\bibnamefont {Zhang}},
  \bibinfo {author} {\bibfnamefont {W.}~\bibnamefont {Walden-Newman}}, \bibinfo
  {author} {\bibfnamefont {X.}~\bibnamefont {Wang}}, \bibinfo {author}
  {\bibfnamefont {J.}~\bibnamefont {Hone}}, \bibinfo {author} {\bibfnamefont
  {C.~W.}\ \bibnamefont {Wong}}, \ and\ \bibinfo {author} {\bibfnamefont
  {S.}~\bibnamefont {Strauf}},\ }\href {http://dx.doi.org/10.1038/ncomms3152}
  {\bibfield  {journal} {\bibinfo  {journal} {Nat Commun}\ }\textbf {\bibinfo
  {volume} {4}},\ \bibinfo {pages} {2152} (\bibinfo {year} {2013})}\BibitemShut
  {NoStop}%
\bibitem [{\citenamefont {Galland}\ \emph {et~al.}(2008)\citenamefont
  {Galland}, \citenamefont {H\"ogele}, \citenamefont {T\"ureci},\ and\
  \citenamefont {Imamo\u{g}lu}}]{Galland2008}%
  \BibitemOpen
  \bibfield  {author} {\bibinfo {author} {\bibfnamefont {C.}~\bibnamefont
  {Galland}}, \bibinfo {author} {\bibfnamefont {A.}~\bibnamefont {H\"ogele}},
  \bibinfo {author} {\bibfnamefont {H.~E.}\ \bibnamefont {T\"ureci}}, \ and\
  \bibinfo {author} {\bibfnamefont {A.}~\bibnamefont {Imamo\u{g}lu}},\ }\href
  {\doibase 10.1103/PhysRevLett.101.067402} {\bibfield  {journal} {\bibinfo
  {journal} {Phys. Rev. Lett.}\ }\textbf {\bibinfo {volume} {101}},\ \bibinfo
  {pages} {067402} (\bibinfo {year} {2008})}\BibitemShut {NoStop}%
\bibitem [{\citenamefont {Lefebvre}\ \emph {et~al.}(2004)\citenamefont
  {Lefebvre}, \citenamefont {Finnie},\ and\ \citenamefont
  {Homma}}]{Lefebvre2004}%
  \BibitemOpen
  \bibfield  {author} {\bibinfo {author} {\bibfnamefont {J.}~\bibnamefont
  {Lefebvre}}, \bibinfo {author} {\bibfnamefont {P.}~\bibnamefont {Finnie}}, \
  and\ \bibinfo {author} {\bibfnamefont {Y.}~\bibnamefont {Homma}},\ }\href
  {http://link.aps.org/doi/10.1103/PhysRevB.70.045419} {\bibfield  {journal}
  {\bibinfo  {journal} {Phys. Rev. B}\ }\textbf {\bibinfo {volume} {70}},\
  \bibinfo {pages} {045419} (\bibinfo {year} {2004})}\BibitemShut {NoStop}%
\bibitem [{\citenamefont {Htoon}\ \emph {et~al.}(2004)\citenamefont {Htoon},
  \citenamefont {O'Connell}, \citenamefont {Cox}, \citenamefont {Doorn},\ and\
  \citenamefont {Klimov}}]{Htoon2004}%
  \BibitemOpen
  \bibfield  {author} {\bibinfo {author} {\bibfnamefont {H.}~\bibnamefont
  {Htoon}}, \bibinfo {author} {\bibfnamefont {M.~J.}\ \bibnamefont
  {O'Connell}}, \bibinfo {author} {\bibfnamefont {P.~J.}\ \bibnamefont {Cox}},
  \bibinfo {author} {\bibfnamefont {S.~K.}\ \bibnamefont {Doorn}}, \ and\
  \bibinfo {author} {\bibfnamefont {V.~I.}\ \bibnamefont {Klimov}},\ }\href
  {\doibase 10.1103/PhysRevLett.93.027401} {\bibfield  {journal} {\bibinfo
  {journal} {Phys. Rev. Lett.}\ }\textbf {\bibinfo {volume} {93}},\ \bibinfo
  {pages} {027401} (\bibinfo {year} {2004})}\BibitemShut {NoStop}%
\bibitem [{\citenamefont {Krummheuer}\ \emph {et~al.}(2002)\citenamefont
  {Krummheuer}, \citenamefont {Axt},\ and\ \citenamefont
  {Kuhn}}]{Krummheuer2002}%
  \BibitemOpen
  \bibfield  {author} {\bibinfo {author} {\bibfnamefont {B.}~\bibnamefont
  {Krummheuer}}, \bibinfo {author} {\bibfnamefont {V.~M.}\ \bibnamefont {Axt}},
  \ and\ \bibinfo {author} {\bibfnamefont {T.}~\bibnamefont {Kuhn}},\ }\href
  {\doibase 10.1103/PhysRevB.65.195313} {\bibfield  {journal} {\bibinfo
  {journal} {Phys. Rev. B}\ }\textbf {\bibinfo {volume} {65}},\ \bibinfo
  {pages} {195313} (\bibinfo {year} {2002})}\BibitemShut {NoStop}%
\bibitem [{\citenamefont {Besombes}\ \emph {et~al.}(2001)\citenamefont
  {Besombes}, \citenamefont {Kheng}, \citenamefont {Marsal},\ and\
  \citenamefont {Mariette}}]{Besombes2001}%
  \BibitemOpen
  \bibfield  {author} {\bibinfo {author} {\bibfnamefont {L.}~\bibnamefont
  {Besombes}}, \bibinfo {author} {\bibfnamefont {K.}~\bibnamefont {Kheng}},
  \bibinfo {author} {\bibfnamefont {L.}~\bibnamefont {Marsal}}, \ and\ \bibinfo
  {author} {\bibfnamefont {H.}~\bibnamefont {Mariette}},\ }\href@noop {}
  {\bibfield  {journal} {\bibinfo  {journal} {Phys. Rev. B}\ }\textbf {\bibinfo
  {volume} {63}},\ \bibinfo {pages} {155307} (\bibinfo {year}
  {2001})}\BibitemShut {NoStop}%
\bibitem [{\citenamefont {Favero}\ \emph {et~al.}(2003)\citenamefont {Favero},
  \citenamefont {Cassabois}, \citenamefont {Ferreira}, \citenamefont {Darson},
  \citenamefont {Voisin}, \citenamefont {Tignon}, \citenamefont {Delalande},
  \citenamefont {Bastard}, \citenamefont {Roussignol},\ and\ \citenamefont
  {Gerard}}]{Favero2003}%
  \BibitemOpen
  \bibfield  {author} {\bibinfo {author} {\bibfnamefont {I.}~\bibnamefont
  {Favero}}, \bibinfo {author} {\bibfnamefont {G.}~\bibnamefont {Cassabois}},
  \bibinfo {author} {\bibfnamefont {R.}~\bibnamefont {Ferreira}}, \bibinfo
  {author} {\bibfnamefont {D.}~\bibnamefont {Darson}}, \bibinfo {author}
  {\bibfnamefont {C.}~\bibnamefont {Voisin}}, \bibinfo {author} {\bibfnamefont
  {J.}~\bibnamefont {Tignon}}, \bibinfo {author} {\bibfnamefont
  {C.}~\bibnamefont {Delalande}}, \bibinfo {author} {\bibfnamefont
  {G.}~\bibnamefont {Bastard}}, \bibinfo {author} {\bibfnamefont
  {P.}~\bibnamefont {Roussignol}}, \ and\ \bibinfo {author} {\bibfnamefont
  {J.~M.}\ \bibnamefont {Gerard}},\ }\href {\doibase
  10.1103/PhysRevB.68.233301} {\bibfield  {journal} {\bibinfo  {journal} {Phys.
  Rev. B}\ }\textbf {\bibinfo {volume} {68}},\ \bibinfo {pages} {233301}
  (\bibinfo {year} {2003})}\BibitemShut {NoStop}%
\bibitem [{sup()}]{supmat}%
  \BibitemOpen
  \href@noop {} {\bibinfo  {journal} {See Supplemental Material for details}\
  }\BibitemShut {NoStop}%
\bibitem [{\citenamefont {Suzuura}\ and\ \citenamefont
  {Ando}(2002)}]{Suzuura2002}%
  \BibitemOpen
\bibfield  {journal} {  }\bibfield  {author} {\bibinfo {author} {\bibfnamefont
  {H.}~\bibnamefont {Suzuura}}\ and\ \bibinfo {author} {\bibfnamefont
  {T.}~\bibnamefont {Ando}},\ }\href {\doibase 10.1103/PhysRevB.65.235412}
  {\bibfield  {journal} {\bibinfo  {journal} {Phys. Rev. B}\ }\textbf {\bibinfo
  {volume} {65}},\ \bibinfo {pages} {235412} (\bibinfo {year}
  {2002})}\BibitemShut {NoStop}%
\bibitem [{\citenamefont {Pennington}\ and\ \citenamefont
  {Goldsman}(2005)}]{Pennington2005}%
  \BibitemOpen
  \bibfield  {author} {\bibinfo {author} {\bibfnamefont {G.}~\bibnamefont
  {Pennington}}\ and\ \bibinfo {author} {\bibfnamefont {N.}~\bibnamefont
  {Goldsman}},\ }\href@noop {} {\bibfield  {journal} {\bibinfo  {journal}
  {Phys. Rev. B}\ }\textbf {\bibinfo {volume} {71}},\ \bibinfo {pages} {205318}
  (\bibinfo {year} {2005})}\BibitemShut {NoStop}%
\bibitem [{\citenamefont {Nguyen}\ \emph {et~al.}(2011)\citenamefont {Nguyen},
  \citenamefont {Voisin}, \citenamefont {Roussignol}, \citenamefont {Roquelet},
  \citenamefont {Lauret},\ and\ \citenamefont {Cassabois}}]{Nguyen2011a}%
  \BibitemOpen
  \bibfield  {author} {\bibinfo {author} {\bibfnamefont {D.~T.}\ \bibnamefont
  {Nguyen}}, \bibinfo {author} {\bibfnamefont {C.}~\bibnamefont {Voisin}},
  \bibinfo {author} {\bibfnamefont {P.}~\bibnamefont {Roussignol}}, \bibinfo
  {author} {\bibfnamefont {C.}~\bibnamefont {Roquelet}}, \bibinfo {author}
  {\bibfnamefont {J.~S.}\ \bibnamefont {Lauret}}, \ and\ \bibinfo {author}
  {\bibfnamefont {G.}~\bibnamefont {Cassabois}},\ }\href {\doibase
  10.1103/PhysRevB.84.115463} {\bibfield  {journal} {\bibinfo  {journal} {Phys.
  Rev. B}\ }\textbf {\bibinfo {volume} {84}},\ \bibinfo {pages} {115463}
  (\bibinfo {year} {2011})}\BibitemShut {NoStop}%
\bibitem [{Note1()}]{Note1}%
  \BibitemOpen
  \bibinfo {note} {Ultra-narrow PL spectra of suspended nanotubes at 4 and 77 K
  provided by A. Hoegele in a private communication could be quantitatively
  fitted to our model with an acoustic barrier width of 80 nm and height of 1
  meV}\BibitemShut {NoStop}%
\bibitem [{Note2()}]{Note2}%
  \BibitemOpen
  \bibinfo {note} {In some other cases where no anti-bunching signature is
  observed \cite {Sarpkaya2013}, the exciton could be more delocalized, which
  would also suppress the wings.}\BibitemShut {Stop}%
\bibitem [{Note3()}]{Note3}%
  \BibitemOpen
  \bibinfo {note} {We note that self trapping of the exciton with acoustic
  phonon can also lead to a modification of the phonon spectrum in the meV
  range \cite {Shaw1981}}\BibitemShut {NoStop}%
\bibitem [{\citenamefont {Crochet}\ \emph {et~al.}(2012)\citenamefont
  {Crochet}, \citenamefont {Duque}, \citenamefont {Werner}, \citenamefont
  {Lounis}, \citenamefont {Cognet},\ and\ \citenamefont {Doorn}}]{Crochet2012}%
  \BibitemOpen
  \bibfield  {author} {\bibinfo {author} {\bibfnamefont {J.~J.}\ \bibnamefont
  {Crochet}}, \bibinfo {author} {\bibfnamefont {J.~G.}\ \bibnamefont {Duque}},
  \bibinfo {author} {\bibfnamefont {J.~H.}\ \bibnamefont {Werner}}, \bibinfo
  {author} {\bibfnamefont {B.}~\bibnamefont {Lounis}}, \bibinfo {author}
  {\bibfnamefont {L.}~\bibnamefont {Cognet}}, \ and\ \bibinfo {author}
  {\bibfnamefont {S.~K.}\ \bibnamefont {Doorn}},\ }\bibfield  {booktitle}
  {\emph {\bibinfo {booktitle} {Nano Letters}},\ }\href
  {http://dx.doi.org/10.1021/nl301739d} {\bibfield  {journal} {\bibinfo
  {journal} {Nano Lett.}\ }\textbf {\bibinfo {volume} {12}},\ \bibinfo {pages}
  {5091} (\bibinfo {year} {2012})}\BibitemShut {NoStop}%
\bibitem [{\citenamefont {Savin}\ \emph {et~al.}(2009)\citenamefont {Savin},
  \citenamefont {Hu},\ and\ \citenamefont {Kivshar}}]{Savin2009}%
  \BibitemOpen
  \bibfield  {author} {\bibinfo {author} {\bibfnamefont {A.~V.}\ \bibnamefont
  {Savin}}, \bibinfo {author} {\bibfnamefont {B.}~\bibnamefont {Hu}}, \ and\
  \bibinfo {author} {\bibfnamefont {Y.~S.}\ \bibnamefont {Kivshar}},\ }\href
  {\doibase 10.1103/PhysRevB.80.195423} {\bibfield  {journal} {\bibinfo
  {journal} {Phys. Rev. B}\ }\textbf {\bibinfo {volume} {80}},\ \bibinfo
  {pages} {195423} (\bibinfo {year} {2009})}\BibitemShut {NoStop}%
\bibitem [{\citenamefont {Shaw}\ and\ \citenamefont {Young}(1981)}]{Shaw1981}%
  \BibitemOpen
  \bibfield  {author} {\bibinfo {author} {\bibfnamefont {P.~B.}\ \bibnamefont
  {Shaw}}\ and\ \bibinfo {author} {\bibfnamefont {E.~W.}\ \bibnamefont
  {Young}},\ }\href {\doibase 10.1103/PhysRevB.24.714} {\bibfield  {journal}
  {\bibinfo  {journal} {Phys. Rev. B}\ }\textbf {\bibinfo {volume} {24}},\
  \bibinfo {pages} {714} (\bibinfo {year} {1981})}\BibitemShut {NoStop}%
\end{thebibliography}%

\end{document}